\documentstyle[aps,pra,epsfig]{revtex}

\begin{document}

\draft

\title{Information theoretic aspects in ponderomotive systems}

\author{Silvia Giannini, Stefano Mancini and Paolo Tombesi}

\address{INFM, Dipartimento di Fisica, Universit\`a di Camerino,
I-62032 Camerino, Italy
}

\date{\today}

\maketitle

\begin{abstract}
We show the possibility to entangle radiation modes
through a simple reflection on a moving mirror.
The model of an optical cavity having a movable end mirror,
and supporting different modes is employed.
The mechanical motion of the mirror mediates information 
between the modes leading to an effective mode-mode
interaction. We characterize the modes' entanglement 
on the basis of recent separability criteria. 
The effect of the thermal
noise associate to the mirror's motion is accounted for.
Then, we evaluate the performances of such 
{\it ponderomotive entanglement} in possible applications 
like teleportation and telecloning.
\end{abstract}


\section{Introduction}

Ponderomotive systems are physical systems
where the electromagnetic 
pressure force gives rise to relevant effects. 
The optomechanical coupling between
a movable mirror and a radiation field,
is realized in such systems
when the field is reflected by the moving mirror. 
This coupling was introduced in the context of quantum limited 
measurements  \cite{BKbook} and then used in interferometric 
gravitational-wave detection \cite{ABR} as well as in atomic 
force microscope \cite{AFM}.
Since then, a wide literature has been devoted to such a
coupling. In particular, it has been shown 
that it may lead to
nonclassical states of both the radiation 
field \cite{FAB,PRA94}, 
and the motion of the mirror 
\cite{PRA97}.
The interest about ponderomotive systems 
also relies on the possibility to investigate, with them,
the tricky borderline between the quantum and the classical 
world \cite{SOU,PRL}.
Moreover, recent technical progresses have 
made this area experimentally accessible 
\cite{EXP1,EXP2}.

The appearance of quantum effects in ponderomotive systems,
paves the way to use them also for quantum information purposes
\cite{QI}. These require as main ingredient the entanglement
\cite{SCH,EPR}. 
Furthermore, information processing,
in the quantum optical framework, 
can be implemented when applied to continuous quadratures
of electromagnetic modes \cite{CVbook}.
Then, the use of a ponderomotive meter for continuous 
variable entanglement purification has 
been investigated in Ref.\cite{PLA}. 
Furthermore, the possibility to
obtain quantum correlated quadratures of the field reflected by a 
movable mirror has been proposed in Refs.
\cite{GMT,ALE}.
Here, following the line sketched in Ref.\cite{GMT}, we 
study a ponderomotive system, namely a radiation field 
reflected by an oscillating mirror, 
from the quantum information perspective.
In particular, in Section II, we shall show
that the mechanical motion of the mirror mediates information 
between the field modes leading to an effective mode-mode
interaction. 
In Section III, we shall characterize the modes' entanglement 
on the basis of recent criteria \cite{PRL,DUA,SIM}. 
We shall also account for the effect of the thermal
noise associate to the mirror's motion.
Then, in Section IV, we shall evaluate the performances of such 
ponderomotive entanglement in possible applications 
like teleportation \cite{VAI,BK} and telecloning \cite{VLB}.
Finally, Section V is for concluding remarks.

\section{A ponderomotive system}

\begin{figure}[t]
\centerline{\epsfig{figure=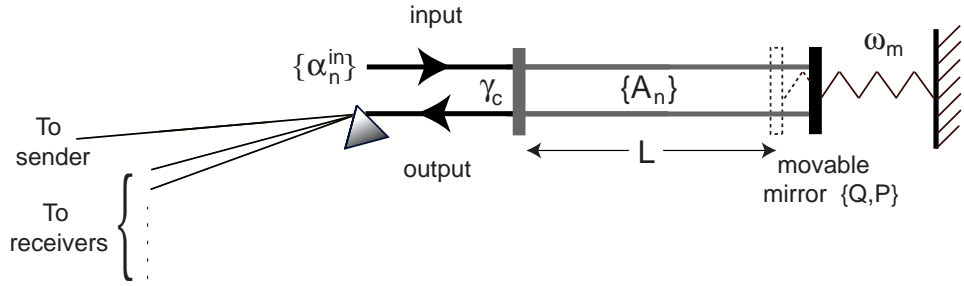,width=5 in}}
\caption{\widetext
A possible scheme implementing the studied ponderomotive system.
}
\label{fig1}
\end{figure}

The model we are going to consider is schematically depicted in 
Fig.\ref{fig1}. It consists of a linear cavity, 
with an oscillating end mirror, plunged in a thermal reservoir
at the equilibrium temperature $T$.
This completely reflecting mirror, with mass $m$, can move back and forth 
along the cavity axes.
When the cavity is empty the moving mirror undergoes harmonic 
oscillations at frequency $\omega_{m}$,
damped at rate $\gamma_{m}$ by the coupling to the external bath. 
In presence of a radiation field,  
the cavity length varies under the action of the radiation pressure 
force, which causes the instantaneous displacement of the mirror.
The resonant frequencies of the cavity are calculated at the 
equilibrium position of the oscillating mirror, resulting 
\begin{equation}\label{wcn}
    \omega_{cn}=\frac{\pi\,c}{L}\,{\tilde n}\,,
\end{equation}    
where ${\tilde n}$ is an arbitrary integer number 
corresponding to the index $n$,
${c}$ is the speed of 
light, and $L$ is the equilibrium cavity length.
We consider the possibility to have several 
input fields at frequencies
${\omega_{0\,n}\sim\omega_{c\,n}\,}$ 
driving the corresponding cavity modes.
In the adiabatic limit in which the mirror frequency is much smaller
than the cavity free spectral range $c/(2L)$ we can focus only on the 
driven cavity modes, obtaining the following Hamiltonian
\begin{mathletters}\label{Ham}
\begin{eqnarray}
    H_{tot}&=&H_{free}+H_{drive}+H_{int}\,,\label{Htot}
    \\
    H_{free}&=&\hbar\sum_{n}\omega_{c\,n}A_{n}^{\dag}A_{n}+\frac{P^2}{2m}
    +\frac{1}{2} m\omega_{m}^2{Q}^2\,,\label{Hfree}
    \\
    H_{drive}&=&i\hbar\sqrt{\gamma_{c}}\sum_{n}\left(
    \alpha_{n}^{in}e^{-i\omega_{0\,n}t}A_{n}^{\dag}
	-\alpha_{n}^{in\,*}e^{i\omega_{0\,n}t}A_{n}\right)\,,
    \label{Hdrive}
    \\
    H_{int}&=&-\hbar\sum_{n}\frac{\omega_{c\,n}}{L}A_{n}^{^\dag}
    A_{n}Q\,,\label{Hint}
\end{eqnarray}
\end{mathletters}
where the sum must be intended over the driven modes.
${H_{free}}$ is the Hamiltonian for the free motion
of the mechanical oscillator (moving mirror) having
position $Q$ and momentum $P$, and 
of the cavity modes characterized by the ladder operators 
$A_{n}\,,\,A_{n}^{\dag}$.
Instead, ${H_{drive}}$ describes the input fields, with amplitudes
$\alpha_{n}^{in}$, entering the cavity 
through the fixed mirror
whose partial transmission determines the input-output 
rate $\gamma_{c}$. 
Finally, ${H_{int}}$ represents the ponderomotive interaction 
between the mirror and the radiation fields \cite{LAW}. 
Such interaction is generated by the radiation pressure
induced variation of the cavity length, which 
corresponds to a variation of the frequencies 
(energy levels) through Eq.(\ref{wcn}),
that is
\begin{equation}\label{dwcn}
   \delta\,\omega_{c\,n}\,=\,\frac{\partial\,\omega_{c\,n}}{\partial\,L}\,
   \partial\,L\,=\,-\,\frac{\omega_{c\,n}}{L}\,Q\,,
\end{equation} 
with ${Q=\delta L\ll L}$. Since we shall consider few modes
whose ${\tilde n}\ll c/(2L)$ differ not too much each other,
we can set
$(\omega_{c\,n}/L)\,\simeq\,G\,,\,\forall\,n$, as 
the optomechanical coupling constant.

By using Eqs.(\ref{Ham}), and accounting for the losses and the noises,
we can describe the complete dynamics of the system 
through the following quantum Langevin equations 
\begin{mathletters}\label{NL}
\begin{eqnarray}
   \dot{Q}(t)&=&\,\frac{P(t)}{m}\,,\\
   \dot{P}(t)&=&\,-\,m\,\omega_{m}^2\,Q(t)\,+\,\hbar\,G\,
   \sum_{n}\,A_{n}^{\dagger}(t)\,A_{n}(t)\,-\,2\,\gamma_{m}\,P(t)\,-\,
   \xi(t)\,,\\
   \dot{A}_{n}(t)&=&\,-\,i\,\left(\omega_{c\,n}\,-\,\omega_{0\,n}\right)
   \,A_{n}(t)\,+\,i\,G\,A_{n}(t)\,Q(t)\,+\,\sqrt{\gamma_{c}}\,\alpha_{n}^{in}
   \,-\,\frac{\gamma_{c}}{2}\,
   A_{n}(t)\,+\,\sqrt{\gamma_{c}}\,a_{n}^{in}(t)\,,
\end{eqnarray}
\end{mathletters}
where we have used the replacements $A_{n}(t)\to A_{n}e^{-i\omega_{0\,n}t}$.
Furthermore, ${a_{n}^{in}}$ are the vacuum noise operators associated 
to the input radiation fields, while ${\xi(t)}$ is the noise operator 
for the quantum Brownian motion of the mirror. The noise correlations are
\cite{GAR,GV}
\begin{mathletters}\label{NCT}
\begin{eqnarray}
   \langle\,a_{j}^{in}(t)\,a_{k}^{in\,\dagger}(t')\,\rangle&&=
   \,\delta\,(t-t^{\prime})\,
   \delta_{j,k}\,,\label{corrt}
   \\
   \langle\,\xi(t)\,\xi(t^{\prime})\,\rangle&&=\,\frac{m\,\gamma_{m}\,
   \hbar}{\pi}\,\int d\omega\,\left\{\frac{\omega\,\left[\,\coth\,\left(\,
   \frac{\hbar\,\omega}{2\,K_{B}\,T}\,\right)\,-\,1\,\right]}
   {\exp\,\left(\,-\,i\,\omega\,\left(t-t^{\prime}\,\right)\,\right)}
   \right\}\,,\label{corrnoiset}
\end{eqnarray}
\end{mathletters}
where $k_{B}$ is the Boltzmann constant. It is worth noting that 
Eq.(\ref{corrnoiset}) gives the exact thermal noise correlations at 
any temperature ${T}$ \cite{GV}.

We are now going to study the dynamics of the small fluctuations
around the steady state, i.e. the dynamics of the operators
\begin{mathletters}
\begin{eqnarray}
	q(t)&=&Q(t)\,-\,x\,,
    \\
    p(t)&=&P(t)\,-\,y\,,
    \\
	a_{n}(t)&=&A_{n}\,-\,\alpha_{n}\,,
\end{eqnarray}
\end{mathletters}
where the (classical) stationary values are given by
\begin{mathletters}\label{ss}
\begin{eqnarray}
    x\,&\equiv&\,\langle\,Q\,\rangle_{ss}\,
	=\,\frac{2\hbar\,G}{m\,\omega_{m}^2}\,
    \sum_{n}\,\mid\alpha_{n}\mid^2\,,\label{x}
    \\
    y\,&\equiv&\,\langle\,P\,\rangle_{ss}\,=\,0\,,\label{y}
    \\
    \alpha_{n}\,&\equiv&\,\langle\,A_{n}\,\rangle_{ss}\,
	=\,\frac{\alpha_{n}^{in}}
    {\sqrt{\gamma_{c}}\,\left(\,\frac{1}{2}\,
	-\,i\,\Delta_{n}\,\right)}\,,\label{alfa}
\end{eqnarray}
\end{mathletters}
with
\begin{equation}
	\Delta_{n}\,=\,\frac{\,\omega_{0\,n}\,
	-\omega_{c\,n}\,+\,G\,x\,}{\gamma_{c}}\,,\label{delta}
\end{equation}
the (dimensionless) 
overall detuning due to the frequency mismatch and to
the radiation phase shift caused by the stationary displacement $x$ 
of the mirror.

For the sake of simplicity we assume,
from now on, symmetric conditions for the 
various radiation modes, that is, 
$\Delta_{n}=\Delta$ and $\alpha_{n}=\alpha \in {\Re}$,
$\forall n$. Then, it is easily recognizable in Eq.(\ref{alfa})
the nonlinear relation between
input and intracavity intensity field
which give rise to the bistable behavior of the system
\cite{DOR}.

Linearizing Eqs.(\ref{NL}) around the steady state 
(\ref{ss}) we obtain
\begin{mathletters}\label{L}
\begin{eqnarray}
    \dot{q}(t)&=&\,\frac{p(t)}{m}\,,
    \\
    \dot{p}(t)&=&\,-\,m\,\omega_{m}^2\,q(t)
    \,+\,\hbar\,G\,\sum_{n}\,
    \left[\,\alpha^*\,a_{n}(t)\,+\,\alpha\,a_{n}^{\dagger}(t)\,
    \right]    
    \,-\,2\,\gamma_{m}\,p(t)\,-
    \,\xi(t)\,,
    \\
    \dot{a}_{n}(t)&=&\,\left(\,i\,\Delta\,-\,\frac{1}{2}\,\right)\, 
    \gamma_{c}\,a_{n}(t)\,+\,i\,G\,\alpha\,q(t)\,+\,\sqrt{\gamma_{c}}\,
    a_{n}^{in}(t)\,.
\end{eqnarray}
\end{mathletters}
Going into the frequency domain, and eliminating the mirror's variables
we are left with a set of $2N$ linear equations 
($N$ being the number of driven modes, so that $n=1,\ldots ,N$)
for the modes quadratures
\begin{mathletters}\label{QUAD}
\begin{eqnarray}
    X_{n}(\omega)&=&\,\frac{
    a_{n}(\omega)\,+\,a_{n}^{\dag}(-\omega)}{\sqrt{2}}\,,
    \\
    Y_{n}(\omega)&=&\,-\,i\,\frac{\,a_{n}(\omega)\,-
    \,a_{n}^{\dag}(-\omega)\,}{\sqrt{2}}\,.\label{opw}
\end{eqnarray}
\end{mathletters}
Such equations can be written in compact form as
\begin{equation}\label{compact}
    i\,\omega\,{\bf v}(\omega)\,=\,{\cal M}(\omega)\,{\bf v}(\omega)\,+
    \,\sqrt{\gamma_{c}}\,{\bf v}^{\,in}(\omega)\,+\,{\bf s}(\omega)\,
    \xi(\omega)\,,
\end{equation}
where we have introduced the $2N$ dimensional vectors
\begin{mathletters}
\begin{eqnarray}
    {\bf v}(\omega)&=&\,\left(\,X_{1}(\omega)\,,\,Y_{1}(\omega)\,,\ldots\,,\,
    X_{N}(\omega)\,,\,Y_{N}(\omega)\,\right)^{T}\,,
    \\
    {\bf v}^{\,in}(\omega)&=&\,\left(\,X_{1}^{in}(\omega)\,,\,Y_{1}^{in}
    (\omega)\,,\ldots\,,\,X_{N}^{in}(\omega)\,,\,Y_{N}^{in}(\omega)\,
    \right)^{T}\,,
    \\
    {\bf s}(\omega)&=&\,\sqrt{2}\,G\,\chi(\omega)\,\left(\,0\,,\,
    -\,\alpha\,,\ldots\,
    0\,,\,-\,\alpha\,
    \right)^{T}\,,\label{vet}
\end{eqnarray}  
\end{mathletters}
with
\begin{equation}\label{ff}
   \chi(\omega)=\,\frac{1}{m\,(\omega_{m}^2\,-\,\omega^2\,+\,2\,i\,
   \gamma_{m}\,\omega)}\,,
\end{equation}
the mirror's mechanical response function.
Furthermore, ${\cal M}(\omega)$ is a ${2N\times 2N}$ matrix 
written as
\begin{equation}\label{M}
    {\cal M}=\left(
    \begin{array}{cccc}
    {\cal M}_{d}&{\cal M}_{o}&\cdots&{\cal M}_{o}
    \\
    {\cal M}_{o}&{\cal M}_{d}&\cdots&{\cal M}_{o}
    \\
    \vdots & 
    \vdots &
    \ddots &
    \vdots 
    \\
    {\cal M}_{o}&{\cal M}_{o}&\cdots&{\cal M}_{d}
    \end{array}
    \right)\,,
\end{equation}
where ${\cal M}_{d}$ and ${\cal M}_{o}$
are $2\times 2$ matrices given by
\begin{mathletters}\label{MdMo}
\begin{eqnarray}
    {\cal M}_{d}&=&\left(
    \begin{array}{cc}
    -\gamma_{c}/2&-\Delta\gamma_{c}
    \\
    \Delta\gamma_{c}+2\hbar G^{2}\chi(\omega)\alpha^{2}&
    -\gamma_{c}/2
    \end{array}
    \right)\,,
    \\
    {\cal M}_{o}&=&\left(
    \begin{array}{cc}
    0&0
    \\
    2\hbar G^{2}\chi(\omega)\alpha^{2}&0
    \end{array}
    \right)\,.
\end{eqnarray}
\end{mathletters}

The useful noise correlations for Eq.(\ref{compact})
come from Eqs.(\ref{NCT}), (\ref{QUAD}) and read 
\begin{mathletters}\label{corrXY}
\begin{eqnarray}
    \left<\,X_{j}^{in}(\omega)\,
    X_{k}^{in}(\omega^{\prime}\,)\,\right>\,&=&\frac{1}{2}\,
    \delta_{j,k}\,\delta(\omega\,+\,\omega^{\prime}\,)\,,
    \\
    \left<\,Y_{j}^{in}(\omega)\,
    Y_{k}^{in}(\omega^{\prime}\,)\,\right>\,&=&\frac{1}{2}\,
    \delta_{j,k}\,\delta(\omega\,+\,\omega^{\prime}\,)\,,
    \\
    \left<\,X_{j}^{in}(\omega)\,
    Y_{k}^{in}(\omega^{\prime}\,)\,\right>\,&=&\frac{1}{2}\,
    i\,\delta_{j,k}\,\delta(\omega\,+\,\omega^{\prime}\,)\,,
\end{eqnarray}
\end{mathletters}
and 
\begin{equation}\label{corrNoise}
    \left<\,\xi(\omega)\,\xi(\omega^{\prime}\,)\,\right>\,=\,
    \left\{\,1\,+\,\coth \,\left(\,
    \frac{\hbar\,\omega}{2\,K_{B}\,T}\,\right)\,\right\}\,
    \frac{m\,\gamma_{m}\,\hbar}{\pi}\,\omega\,
    \delta(\omega\,+\,\omega^{\prime}\,)\,.
\end{equation}
Thus Eqs.(\ref{compact})-(\ref{corrNoise}) completely describe the 
dynamics of the small fluctuations of radiation modes.   
Practically, we can see from Eqs.(\ref{L}) that the mirror
mediates information between the radiation modes leading to an 
effective mode-mode interaction as results from Eqs.(\ref{compact}),
(\ref{M}) and (\ref{MdMo}).

\section{Output fields entanglement}

The above discussed mode-mode interaction will presumably lead
to entanglement between intracavity modes, which, in turn,
should be reflected on the fields outgoing the cavity. 
On the other hand, only these latter become really useful. Hence, we are going 
to characterize their correlations. First of all we notice, by the
the input-output theory \cite{GAR}, that 
\begin{equation}\label{vout}
    {\bf v}^{\,out}(\omega)\,=\,
    \sqrt{\gamma_{c}}\,{\bf v}(\omega)\,-\,
    {\bf v}^{\,in}(\omega)\,,
\end{equation}
then, we introduce the hermitian output quadrature operators 
\begin{mathletters}\label{RXY}
\begin{eqnarray}
   R_{X_{n}^{out}}\,&=&\,\frac{X_{n}^{out}(\omega)\,
   +\,X_{n}^{out}(-\omega)}{2}\,,\nonumber
   \\
   R_{Y_{n}^{out}}\,&=&\,\frac{Y_{n}^{out}(\omega)\,
   +\,Y_{n}^{out}(-\omega)}{2}\,.
\end{eqnarray}
\end{mathletters}
Their correlations are described by the $2N\times 2N$ matrix
\begin{eqnarray}\label{cormat}
    {{\cal G}}&\equiv&\,\frac{1}{4}\,\langle\,{\bf v}^{\,out}(\omega)\,
    \left[{\bf v}^{\,out}(-\omega)\right]^{T}\,
    +\,{\bf v}^{\,out}(-\omega)\,
    \left[{\bf v}^{\,out}(\omega)\right]^{T}\,\rangle\,,
    \\
    &=&\frac{1}{4}\,{\cal K}(\omega)\,\langle\,{\bf  v}^{\,in}(\omega)\,
    \left[{\bf v}^{\,in}(-\omega)\right]^{T}\,\rangle\,
    \left[{\cal K}(-\omega)\right]^{T}\,
    \nonumber\\
    &+&\,\frac{1}{4}\,{\cal K}(-\omega)\,\langle\,{\bf v}^{\,in}(-\omega)\,
    \left[{\bf v}^{\,in}(\omega)\right]^{T}\,\rangle\,
    \left[{\cal K}(\omega)\right]^{T}\,\nonumber
    \\
    &+&\frac{\gamma_{c}}{4}\,{\cal L}^{-1}(\omega)\,{\bf s}(\omega)\,
    \left[{\bf s}(-\omega)\right]^{T}\,\left[{\cal L}^{-1}(-\omega)
    \right]^{T}\,
    \langle\,\xi(\omega)\,\xi(-\omega)\,\rangle\nonumber\\
    &+&\frac{\gamma_{c}}{4}\,{\cal L}^{-1}(-\omega)\,{\bf s}(-\omega)\,
    \left[{\bf s}
    (\omega)\right]^{T}\,\left[{\cal L}^{-1}(\omega)\right]^{T}\,\langle\,
    \xi(-\omega)\,\xi(\omega)\,\rangle\,,\nonumber
\end{eqnarray}
where
\begin{mathletters}\label{AK}
\begin{eqnarray}
    {\cal K}(\omega)&=&\,\gamma_{c}\,{\cal L}^{-1}(\omega)\,-\,{\cal I}\,,
    \\
    {\cal L}(\omega)&=&\,i\,\omega\,{\cal I}\,-\,{\cal M}(\omega)\,,
\end{eqnarray}
\end{mathletters}
with ${\cal I}$ the identity $2N\times 2N$ matrix.
However, the matrix ${\cal G}$ is not symmetric and, moreover,
it concerns quadratures with frequency dependent commutator, i.e.,
\begin{equation}\label{comm}
   \langle\,\left[\,{R}_{X_{n}^{out}}(\omega)\,,\,
   {R}_{Y_{n}^{out}}(-\omega)\,
   \right]\,\rangle\,=\,i\,c(\omega)\,,\quad \forall n\,,
\end{equation}
with ${c}$ a real positive definite function of frequency ${\omega}$. 
Therefore, to easily apply quantum information arguments
to our system, we 
construct from Eqs.(\ref{cormat}), (\ref{comm}) 
a symmetric correlation matrix
concerning quadratures with canonical commutation relations,
that is
\begin{equation}\label{GAmat}
    {\cal V}_{j,k}(\omega)=\frac{
    {{\cal G}}_{j,k}(\omega)+{{\cal G}}_{k,j}(\omega)}
    {2c(\omega)}\,.
\end{equation}
By virtue of the linearization procedure adopted in Sec.II
we have, for each frequency,
a multivariate Gaussian state completely characterized by 
Eq.(\ref{GAmat}).

\subsection{Bipartite entanglement}

\begin{figure}[t]
\centerline{\epsfig{figure=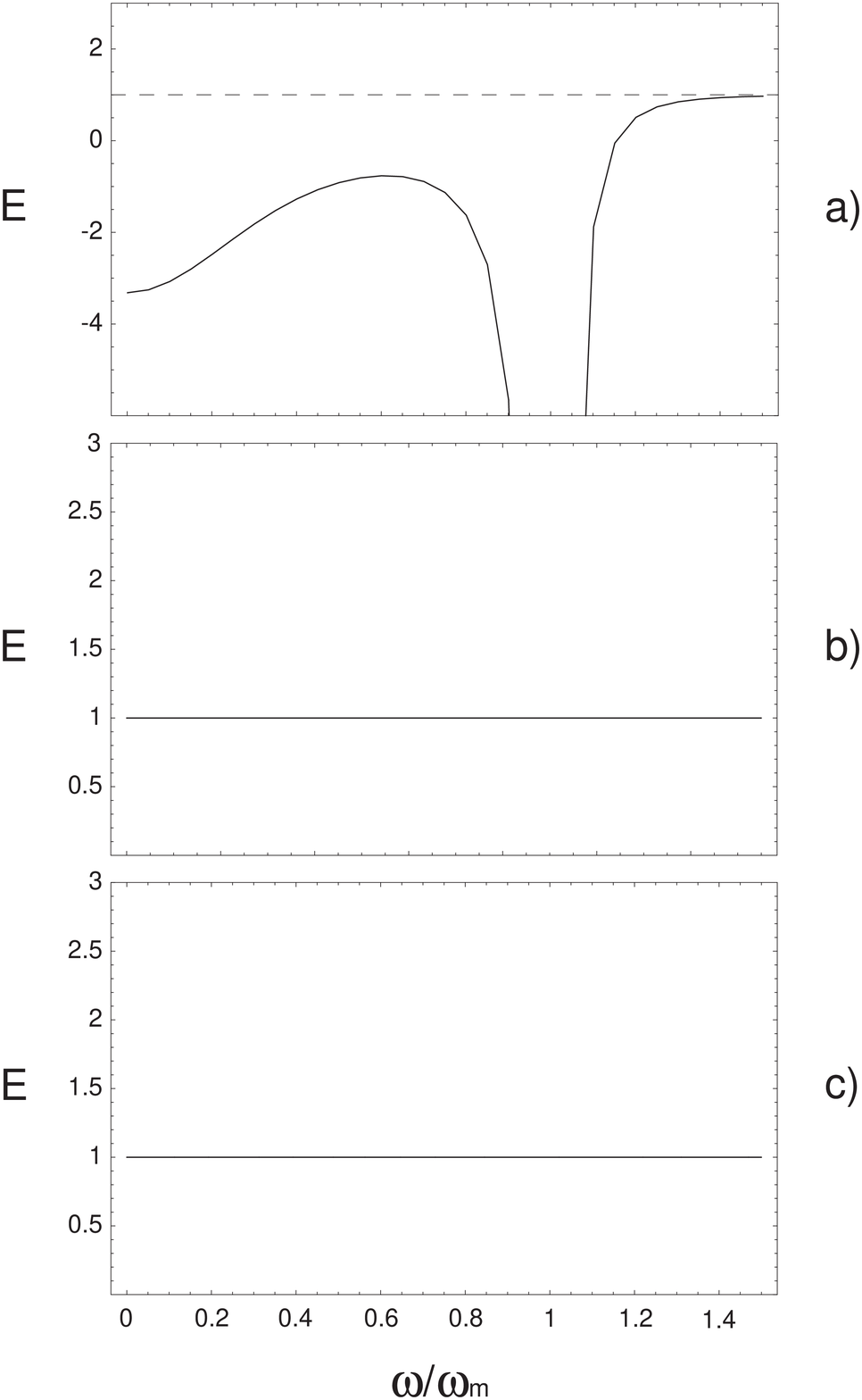,width=3.5in}}
\vspace{0.5cm}
\caption{\widetext
The marker of entanglement $E$ is plotted versus $\omega$
with respect to the three criteria a) Simon, b) product, c) sum. 
Here $\Delta=0$ and the other parameter values are given 
in Tab.I. The dashed lines indicate the limiting value below which 
entanglement is recognized. 
The three plots remain unalterated in the temperature range 
$T=0\,\div\,300$ K. 
}
\label{fig2}
\end{figure}

We now restrict the attention to only two mode ($N=2$)
in order to study the entanglement of a bipartite system.
Practically we consider 
\begin{equation}\label{M2}
    {\cal M}=\left(
    \begin{array}{cc}
    {\cal M}_{d}&{\cal M}_{o}
    \\
    {\cal M}_{o}&{\cal M}_{d}
    \end{array}
    \right)\,,
\end{equation}
and we introduce the matrices 
\begin{equation}
    {\cal J}=\left(
    \begin{array}{cc}
    0&1
    \\
    -1&0
    \end{array}
    \right)\,,
    \qquad
    {\cal R}=\left(
    \begin{array}{cc}
    1&0
    \\
    0&-1
    \end{array}
    \right)\,.
\end{equation}
Then, Eq.(\ref{M2}) leads to
\begin{equation}\label{V2}
    {\cal V}=\left(
    \begin{array}{cc}
    {\cal A}&{\cal C}
    \\
    {\cal C}^{T}&{\cal A}
    \end{array}
    \right)\,,
\end{equation}
where ${\cal A}$ and ${\cal C}$ are $2\times 2$ matrices. 
In this case the Simon's criterion \cite{SIM} is necessary and sufficient
for entanglement, and, according to Eq.(19) of Ref.\cite{SIM}
we can define a marker of entanglement as
\begin{equation}
    E=1+\left({\rm det}{\cal A}\right)^{2}
    +\left(\frac{1}{4}-|{\rm det}{\cal C}|\right)^{2}
    -{\rm tr}\left\{{\cal AJCJAJC}^{T}{\cal J}\right\}
    -\frac{1}{2}{\rm det}{\cal A}\,,
\end{equation}
so that, if it goes below $1$, the state is entangled.
Instead, the product criterion introduced in Ref.\cite{PRL},
and reminiscent of nonlocality criterion \cite{REID},
gives
\begin{equation}
    E=4({\cal A}_{11}+{\cal C}_{11})({\cal A}_{22}-{\cal C}_{22})\,.
\end{equation}
Finally, the sum criterion, expressed by Eq.(3) of Ref.\cite{DUA}, 
can be rewritten as 
\begin{equation}\label{esum}
    E={\rm tr}\{{\cal A}\}+{\rm tr}\left\{{\cal CR}\right\}\,.
\end{equation}
Then, in Fig.\ref{fig2} we report the 
marker of entanglement $E$ versus $\omega$,
for the three criteria a) Simon, b) product, c) sum. The parameters values
 are taken similar to those of the 
experimental set up of Ref.\cite{EXP2}. They are written in Tab.I.
Fig.\ref{fig2} shows that in case of no detuning 
the product and the sum criteria do not reveal any entanglement,
while the Simon criterion do. This proves the weakness of the
the entanglement coming in this case from the 
interaction of only amplitude quadratures
as can be evicted from Eqs.(\ref{M2}) and (\ref{MdMo}).
Such type of entanglement, although resistant to thermal effects,
is practically useless in information processing like teleportation 
whose performance are much more related 
to the product and sum criteria \cite{FUC}.

\begin{table}\label{parametri}
\caption{Parameter values}
\begin{tabular}{cc}
$\omega_{m}$ & $10^6$ ${\rm s}^{-1}$ \\
$\omega_{0\,n}$ & $10^{15}$ ${\rm s}^{-1}$ \\
$m$ & $10^{-4}$ ${\rm kg}$ \\
$L$ & $10^{-3}$ ${\rm m}$ \\
$\gamma_{m}$ & $1$ ${\rm s}^{-1}$ \\ 
$\gamma_{c}$ & $10^6$ ${\rm s}^{-1}$ \\
$P_{n}^{in}\,=\hbar \omega_{0\,n} |\alpha_{n}^{in}|^{2}$ &
$13$ ${\rm mW\,per\,mode}$
\end{tabular}
\end{table}

\begin{figure}[t]
\centerline{\epsfig{figure=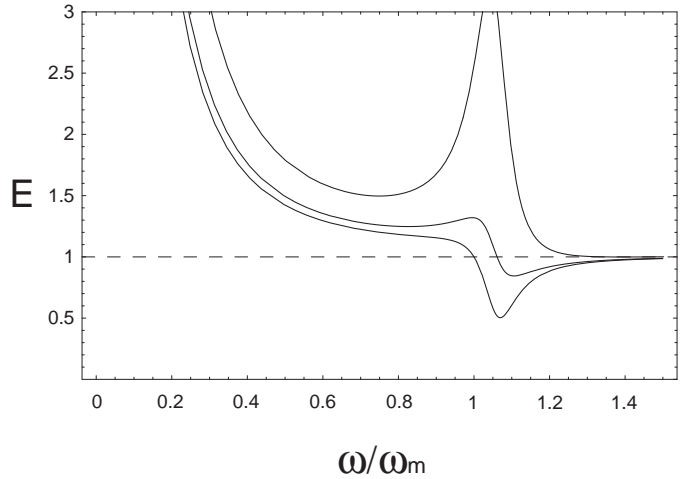,width=3.5in}}
\vspace{0.5cm}
\caption{\widetext
The entanglement indicator $E$ of Eq.(\ref{esum}) is plotted versus $\omega$.
Here $\Delta=-0.1$, and solit lines are for $T=0$, $T=10$, $T=50$ K 
from botton to top. The other parameter values are given 
in Tab.I. The dashed line indicates the limiting value 
below which entanglement is recognized.
}
\label{fig3}
\end{figure}

\begin{figure}[t]
\centerline{\epsfig{figure=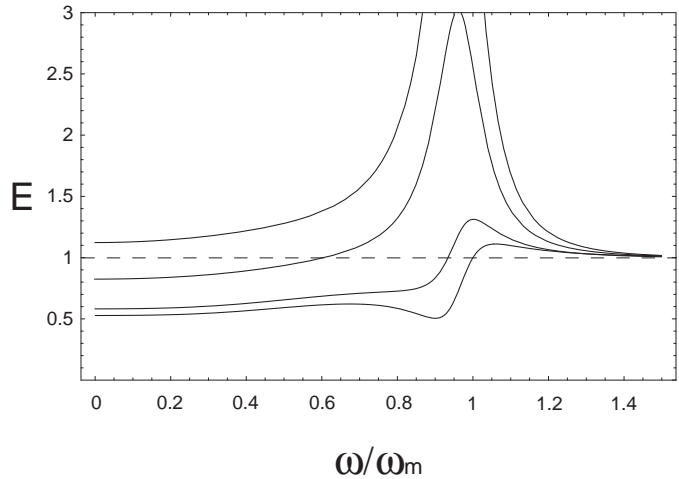,width=3.5in}}
\vspace{0.5cm}
\caption{\widetext
The entanglement indicator $E$ of Eq.(\ref{esum}) is plotted versus $\omega$.
Here $\Delta=0.1$, and solit lines are for $T=0$, $T=10$, $T=50$, 
$T=100$ K from botton to top. The other parameter values are given 
in Tab.I. The dashed line indicates the limiting value 
below which entanglement is recognized.
}
\label{fig4}
\end{figure}

Then in Figs.\ref{fig3} and \ref{fig4} we have only considered the 
sum criterion and we have shown the beneficial effect 
of the detuning on the entanglement. As matter of fact it allows interaction 
of both amplitude and phase quadratures as can be seen in Eq.(\ref{M2}).
In particular Fig.\ref{fig4} exhibits the presence of entanglement at 
low frequencies (according to Ref.\cite{GMT}) as well as near the 
mechanical resonance. Neverthless, the latter turns out to be more 
sensible to the thermal noise. This behavior resembles that of the 
light squeezing studied in Refs.\cite{FAB,PRA94}.

\subsection{Tripartite entanglement}

Characterization of multipartite entanglement ($N>2$) is 
a more complex issue \cite{CVbook}.
In general, multi-party inseparability criteria cannot be formulated 
in compact form as for the two-party. Here, we consider $N=3$, thus the matrix
\begin{equation}\label{M3}
    {\cal M}=\left(
    \begin{array}{cccc}
    {\cal M}_{d}&{\cal M}_{o}&{\cal M}_{o}
    \\
    {\cal M}_{o}&{\cal M}_{d}&{\cal M}_{o}
    \\
    {\cal M}_{o}&{\cal M}_{o}&{\cal M}_{d}
    \end{array}
    \right)\,,
\end{equation}
which leads to 
\begin{equation}\label{V3}
    {\cal V}\,=\,\left(\,
    \begin{array}{ccc}
    {\cal A} & {\cal C} & {\cal C} \\
    {\cal C}^{T} & {\cal A} & {\cal C} \\
    {\cal C}^{T} & {\cal C}^{T} & {\cal A} 
    \end{array}
    \,\right)\,.
\end{equation}
Although for three-mode Gaussian states there exist a necessary and sufficient 
separability criterion \cite{GIE}, its violation does not necessarily 
witness genuine tripartite entanglement. However, from the symmetry 
of matrix (\ref{V3}) we easily deduce that the conditions for 
bipartite entanglement also give tripartite entanglement. 
As matter of fact a  tripartite fully inseparable state is that which 
cannot be separate for any grouping of the parties \cite{GIE}. But, 
due to the symmetry among the parties, if two of them show 
entanglement according to Sec. III A, then any two of them show 
entanglement, thus revealing a fully 
inseparable state.

\section{Applications to remote state transfer}

This ponderomotive entanglement find possible applications 
in quantum information processings with continuous variables 
\cite{CVbook}. Here, we deal with the possibility of using it  
for remote state transfer. By referring to Fig.\ref{fig1},
the modes outgoing the cavity can be separated and one 
of them can reach a sending station while the others 
reach receiving stations.
Then, all these modes constitute the quantum channel
to exploit for transferring a quantum state from the sending
station to the receiving ones.
We will analyze in detail the case for $N=2$, i.e., teleportation 
\cite{VAI,BK},
and $N=3$, i.e., telecloning \cite{VLB}.

\subsection{Teleportation}

The standard teleportation protocol
for continuous variable \cite{VAI,BK} 
can be described by a convolution
of the Wigner functions \cite{CKW}  
\begin{mathletters}\label{Wouttp}
\begin{eqnarray}
    W_{r}(\beta)\,&=&\int d^2\xi\, W_{s}(\xi)\,
    K(\beta-\xi)\,,
    \\
    K(\beta-\xi)&=&\int 
    d^2\xi'\,W({\xi'}^{*}-\xi^{*}\,,\,
    \beta-\xi')\,,
\end{eqnarray}
\end{mathletters}
where ${W_{r}}$ is the Wigner function of the received state,
${W_{s}}$ that of the unknown state to be transferred (sent), 
and ${W}$ that describing the quantum channel between the two stations,
i.e., the two entangled modes characterized by the correlation 
matrix (\ref{V2}).
Here, small greek letters are for complex variables.

By Fourier transforming Eqs.(\ref{Wouttp}), we get a
simple relation
for the characteristic functions $\Phi$, namely 
\begin{equation}\label{chi-out}
    \Phi_{r}(\lambda)\,=\,\Phi_{s}(\lambda)\,
    \tilde{K}(\lambda)\,,
\end{equation}
where
\begin{eqnarray}\label{K-tilde}
    \tilde{K}(\lambda)\,&\equiv& \int d^{2}\kappa\,K(\kappa)\,
    \exp(-i\kappa_{1}\lambda_{1}-i\kappa_{2}\lambda_{2})\,
    \nonumber\\
    &=&\int d^{2}\kappa\,d^{2}\mu\,d^4{\bf z}\, 
    \,\Phi({\bf z}\,)\,\exp\left(-i\kappa_{1}\lambda_{1}
    -i\kappa_{2}\lambda_{2}\right)\nonumber\\
    &\times& \exp\left\{i\,{\bf z}\cdot
    (\mu_{1},-\mu_{2},\kappa_{1}-\mu_{1},\kappa_{2}-\mu_{2})\right\}\,,
\end{eqnarray}
where the variables with the subscript $1$ ($2$) represent
the real (imaginary) part of the corresponding complex variables,
and ${\bf z}$ is a four dimensional real variables vector.
Moreover,
\begin{equation}\label{chi-AB}
    \Phi({\bf z}\,)\,=\,
    \exp\left\{-\frac{1}{4}\,{\bf z}\,\,{\cal V}
    \,{\bf z}^{T}\right\}\,,
\end{equation}
is the characteristic function describing the two-mode channel, thus 
characterized by the matrix ${\cal V}$ given in Eq.(\ref{V2}).

We also consider a Gaussian state to be transferred, so that
\begin{equation}\label{chi-in}
    \Phi_{s}(\lambda)\,=\,\exp\left[\,-\frac{1}{4}\,
    \left(\lambda_{1},\lambda_{2}\right)
    \,{\cal D}\,
    \left(\lambda_{1},\lambda_{2}\right)^{T}\,\right]\,,
\end{equation}
with ${\cal D}$ the $2\times 2$ correlation matrix.

\begin{figure}[t]
\centerline{\epsfig{figure=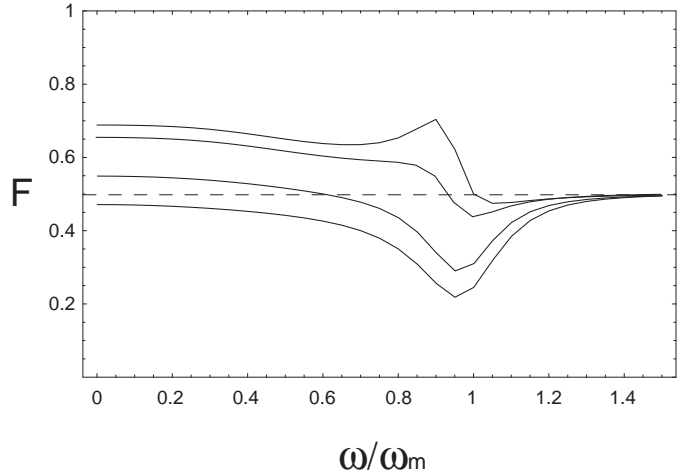,width=3.5in}}
\caption{\widetext
Teleportation fidelity versus $\omega$.
Curves from top to bottom are for $T=0$, $T=10$, $T=50$, $T=100$ K.
Here $\Delta=0.1$ and the values of other parameters 
listed in Tab.I. The dashed line indicates the classical upper bound.
}
\label{fig5}
\end{figure}

Finally, the fidelity of the protocol, 
resulting from the overlap between the ``$r$" and the ``$s$" 
Wigner functions, can be written 
in terms of characteristic functions as
\begin{equation}\label{Fdef}
    F\,\equiv\,\frac{1}{4\,\pi}\int 
    d^{2}\lambda\,\,\Phi_{s}(\lambda)\,\Phi_{r}^{*}(\lambda)\,.
\end{equation}
Then,
by using Eqs.(\ref{chi-out})-(\ref{chi-in}), 
we arrive at (see also \cite{FIU})
\begin{eqnarray}\label{FIDmatrix}
    F\,&=&\,\frac{1}{4\,\pi}\int 
    d^{2}\lambda\,\exp\left[
    -\frac{1}{2}\,(\lambda_{1},\lambda_{2})
    \,{\cal D}\,(\lambda_{1},\lambda_{2})^{T}\right]\,
    \nonumber\\
    &\times&\exp\left[
    -\frac{1}{4}\,
    (\lambda_{1},-\lambda_{2},\lambda_{1},\lambda_{2})
    \,{\cal V}\,
    (\lambda_{1},-\lambda_{2},\lambda_{1},\lambda_{2})^{T}
    \right]\nonumber\\
    &=&\,\frac{1}{\sqrt{\det\left(2{\cal D}+{\cal R}^{T}{\cal AR}
    +{\cal R}^{T}{\cal C}+
    {\cal C}^{T}{\cal R}+{\cal A}\right)}}\,.
\end{eqnarray}
In Fig.\ref{fig5} we show the teleportation fidelity 
as function of $\omega$. As a state to be teleported we have chosen 
the coherent state for which ${\cal D}={\rm diag}(1/2,1/2)$.
In such a case the upper bound for the fidelity achievable 
with only classical means and no quantum resources is $1/2$
\cite{FUC}. Then, we see that this bound is overcame 
just in correspondence of the minima of Fig.\ref{fig3}.
Also the behavior of the fidelity in terms of 
thermal noise reflects that of the 
entanglement in Fig.\ref{fig4}.

\subsection{Telecloning}

As a simple extension of the arguments used for teleportation, we can 
write the Wigner function of the received state (at the two 
stations) by the convolution \cite{VLB}
\begin{mathletters}
\begin{eqnarray}
    {\breve W}_{r}(\beta,\eta)\,&=&\int d^2\xi\, W_{s}(\xi)\,
    {\breve K}(\beta-\xi,\eta-\xi)\,,\label{Wclon}
    \\
    {\breve K}(\beta-\xi,\eta-\xi)&=&
    \int d^2\xi'\,W({\xi'}^{*}-\xi^{*}\,
    ,\,\beta-\xi'\,,\,\eta-\xi')\,,\label{Kclon}
\end{eqnarray}
\end{mathletters}
where $W$ is the Wigner function describing the quantum channel 
between sending and receiving stations, i.e. the three entangled modes
characterized by the correlation matrix (\ref{V3}).

The state at one receiving station can be obtained by tracing the 
received state (\ref{Wclon}) over the other receiving 
station. Due to the symmetry, 
the two possible states coming out coincide. 
Thus, we can assume
\begin{equation}\label{WclonB}
    W_{r}\,(\beta)\,=
    \int d^{2}\eta\,{\breve W}_{r}(\beta,\eta)
    =\int d^{2}\xi W_{s}(\xi) K(\beta-\xi)\,,
\end{equation}
where now
\begin{equation}\label{KclonB}
    K(\beta-\xi)\,=
    \int d^{2}\eta\,{\breve K}(\beta-\xi,\eta-\xi)\,.
\end{equation}
By again Fourier transforming Eq.(\ref{WclonB}), 
we end up with a relation for the characteristic functions
identical to Eq.(\ref{chi-out}), 
\begin{equation}\label{chi-out-tc}
    \Phi_{r}(\lambda)\,=\,\Phi_{s}(\lambda)\,
    \tilde{K}(\lambda)\,,
\end{equation}
where now
\begin{eqnarray}\label{K-tildeclon}
    \tilde{K}(\lambda)\,&=&\int d^{2}\kappa\,
    \exp\left(-i\kappa_{1}\lambda_{1}-i\kappa_{2}\lambda_{2}
    \right)\nonumber\\
    &\times&\int 
    d^{2}\mu\,d^{2}\zeta\,d^6{\bf z}\,\Phi({\bf z}\,)\,
    \exp\left\{i{\bf z}\cdot(\mu_{1},-\mu_{2},\kappa_{1}-\mu_{1},
    \kappa_{2}-\mu_{2},
    \zeta_{1}-\mu_{1},\zeta_{2}-\mu_{2})^{T}\right\}\,,
\end{eqnarray}
and
\begin{equation}\label{chi-ABC}
    \Phi({\bf z})\,=\,\exp\left[-\frac{1}{4}\,{\bf z}\,
    \,{\cal V}\,{\bf z}^{T}
    \right]\,,
\end{equation}
is the characteristic function describing the three-mode channel, thus 
characterized by the matrix ${\cal V}$ given in Eq.(\ref{V3}),
with ${\bf z}$ a 6 dimensional real variables vector.

We again consider a Gaussian state to be transferred,
as in Eq.(\ref{chi-in}). Then, the fidelity,
being expressed by Eq.(\ref{Fdef}), 
results, by means of Eqs.
(\ref{chi-in}), (\ref{chi-out-tc}), (\ref{K-tildeclon}),
(\ref{chi-ABC}) as
\begin{eqnarray}\label{FIDclonB}
    F\,&=&\,\frac{1}{4\,\pi}\int 
    d^{2}\lambda\,\exp\left[
    -\frac{1}{2}\,(\lambda_{1},\lambda_{2})
    \,{\cal D}\,(\lambda_{1},\lambda_{2})^{T}\right]\,
    \nonumber\\
    &\times&\exp\left[
    -\frac{1}{4}\,
    (\lambda_{1},-\lambda_{2},\lambda_{1},\lambda_{2},0,0)
    \,{\cal V}\,
    (\lambda_{1},-\lambda_{2},\lambda_{1},\lambda_{2},0,0)^{T}
    \right]\nonumber\\
    &=&\,\frac{1}{\sqrt{\det\left(2{\cal D}+{\cal R}^{T}{\cal AR}
    +{\cal R}^{T}{\cal C}+
    {\cal C}^{T}{\cal R}+{\cal A}\right)}}\,.
\end{eqnarray}
It practically coincides with Eq.(\ref{FIDmatrix}). However, 
in this case, $F$ is limited above by $2/3$ \cite{CER},
due to the no-cloning theorem \cite{WZ}.

\begin{figure}[t]
\centerline{\epsfig{figure=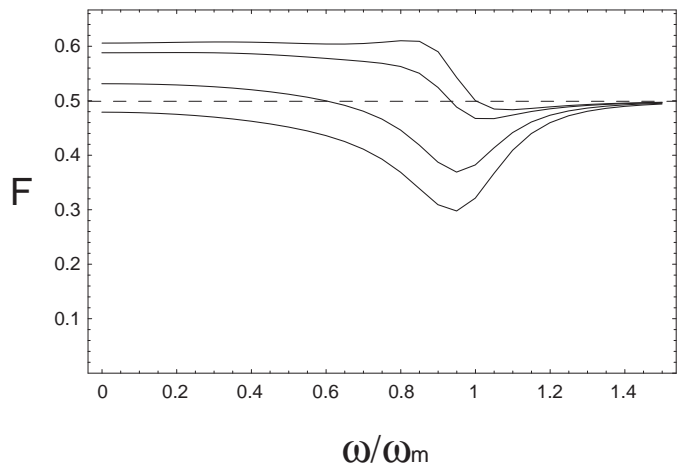,width=3.5in}}
\caption{\widetext
Telecloning fidelity versus $\omega$.
Curves from top to bottom are for $T=0$, $T=10$, $T=50$, $T=100$ K.
Here $\Delta=0.1$ and the values of other parameters 
listed in Tab.I. The dashed line indicates the classical upper bound.
}
\label{fig6}
\end{figure}

In Fig.\ref{fig6} we show the telecloning fidelity 
as function of $\omega$. As state to be telecloned we have choosen
again a coherent state for which ${\cal D}={\rm diag}(1/2,1/2)$.
Also in this case the upper classical bound for the fidelity 
is $1/2$ \cite{CER}. 
Then, we see that this bound is overcome 
again in correspondence of the minima of Fig.\ref{fig4}.
 
\section{Conclusions}

In conclusion, we have studied {\it ponderomotive entanglement},
that is entanglement between radiation modes generated by
the radiation pressure effects. In doing so
we have also provided a comparison between different entanglement 
criteria.
Practically, we have shown that even a classical force,
like radiation pressure force, together with macroscopic objects
can be used for quantum information purposes.
The fidelity achievable in remote state transfer 
widely overcome the classical bounds, thought not
reaching the optimal values. 
Theoretically, this could be obtained
by an optimization of all involved parameters.
However, that would
require large numerical resources without 
adding new physics to the problem. 
So it has been left apart.
Instead, we have investigated the role played by the 
thermal noise related to the mechanical motion of the mirror.
We have seen that purely quantum effects can survive
up to a temperature $\approx 10$ K.
This is within reach in experiments 
with really macroscopic mirrors \cite{EXP2}.
On the other hand, 
the use of micro-opto-mechanical-systems (MOMS)   
\cite{MICRO} surely guarantees better performances.

Beside all that we want to remark that 
ponderomotive systems have a foundational interest \cite{SOU},
and, an information theoretic approach  
could help us in understanding 
the tricky borderline between classical 
and quantum worlds.

\end{document}